\documentclass[12pt]{article}

\usepackage{tikz}
\usetikzlibrary{patterns}
\usepackage{subfigure}
\usepackage[numbers]{natbib}
\usepackage[toc,page]{appendix} 
\usepackage{titling}
\usepackage{amsmath}
\usepackage{tikz}
\usepackage{caption}
\usepackage{algorithm}
\usepackage{algorithmic}
\usepackage[colorlinks = true,linkcolor = blue,citecolor=blue,urlcolor=blue,allbordercolors=white]{hyperref}
\usepackage{doi}
\usepackage{authblk} 
\usepackage[margin=1.0in]{geometry}
\usepackage{graphicx,import}
\usepackage{amsmath}
\usepackage{physics}
\title{\textbf{Network-community analysis of cellular senescence}}
\author[1]{Alda Sabalic}
\author[1]{Victoria Moiseeva}
\author[1,2]{Andres Cisneros}
\author[1]{Oleg Deryagin}
\author[1,2]{Eusebio Perdiguero}
\author[1,2]{Pura Muñoz-Cánoves}
\author[1]{Jordi Garcia-Ojalvo}
\affil[1]{Department of Medical and Life Sciences, Universitat Pompeu Fabra, 08003 Barcelona, Spain}
\affil[2]{Altos Labs, Inc., San Diego Institute of Science, San Diego, CA 92121, USA}

\begin{document}
%
{\let\newpage\relax\maketitle}

\begin{abstract}
Most cellular phenotypes are genetically complex.
Identifying the set of genes that are most closely associated with a specific cellular state is still an open question in many cases.
Here we study the transcriptional profile of cellular senescence using a combination of network-based approaches, which include eigenvector centrality feature selection and community detection.
We apply our method to cell-type-resolved RNA sequencing data obtained from injured muscle tissue in mice.
The analysis identifies some genetic markers consistent with previous findings, and other previously unidentified ones, which are validated with previously published single-cell RNA sequencing data in a different type of tissue.
The key identified genes, both those previously known and the newly identified ones, are transcriptional targets of factors known to be associated with established hallmarks of senescence, and can thus be interpreted as molecular correlates of such hallmarks.
The method proposed here could be applied to any complex cellular phenotype even when only bulk RNA sequencing is available, provided the data is resolved by cell type.
\end{abstract}

\noindent
\textbf{Keywords}: Cellular senescence, molecular hallmarks, RNA sequencing, network centrality, community detection
\section{Introduction}
As cells age, they tend to accumulate damage. While in younger organisms the damage is usually correctly repaired, in older individuals it increasingly leaves traces. One of the direct consequences of this process is cellular senescence, a phenomenon resulting from unresolved molecular damage, which triggers an inflammatory response within the host tissue. 
Senescence is a stable and permanent state of growth arrest in which cells are unable to proliferate \citep{kumari2021}.
It can occur in healthy cells that experience a chronic damage response, involving either direct damaging of the DNA or events like telomere shortening or oncogenic mutations \citep{dekeizer2017}. 
The affected cells go through an irreversible cell cycle arrest by which the effects of the damage become limited \citep{baar2018}. 
Additionally, senescent cells secrete a broad range of growth factors, pro-inflammatory proteins, and matrix proteinases, which alter the cellular microenvironment, in what is known as the senescence-associated secretory phenotype (SASP) \citep{coppe2010}. 

Multiple lines of evidence show that senescent cells are directly implicated in causing age-related phenotypes. 
In studies where senescent cells were genetically or pharmacologically removed, both rapidly and naturally aged mice stayed healthy much longer, and, in some cases, even displayed signs of aging reversal \citep{baker2011, bussian2018}.
The opposite process has also been proven to be true: introducing only a small number of senescent cells into young mice resulted in physical dysfunction \citep{xu2018}. 
Moreover, it was found that senescent cells persist for extended periods of time, which leads to their accumulation during aging \citep{childs2015}.  

Even though senescent cells have shown high expression of cell cycle regulators such as proteins p16 and p21, a universally accepted criterion for their identification is still lacking \citep{cohn2023}. 
Techniques such as RNA sequencing (both bulk and single-cell) can offer valuable insights by comparing the global transcriptional activity of senescent and non-senescent cells.
However, traditional data analysis approaches have not revealed clear transcriptional signatures so far, probably due to the fact that senescence is a highly complex phenotype with multiple cellular implications.
Within that context, the aim of this study was to complement the standard bioinformatics approaches by using a network-based method, which enables us to perform a ``last-mile'' filtering of the set of analyzed genes, and identify a small number of genes that are highly central in distinguishing the senescent from the non-senescent transcriptional profiles.

Complex networks serve as a representation of interactions and connections between their elements --which in the context of this investigation are both genes and cell states. 
Importantly, complex networks are usually characterized by a community structure, meaning that elements belonging to a community are highly interconnected to each other and therefore grouped together.
Conversely, different communities are loosely associated with each other, and their elements are considered less similar.
The community structure that arises from a network constructed using transcriptional data can provide useful information about the transcriptional determinants of a complex cell state such as senescence.

\section{Experimental data}

\begin{figure}[htpb]
\centerline{\includegraphics[width=0.3\linewidth]{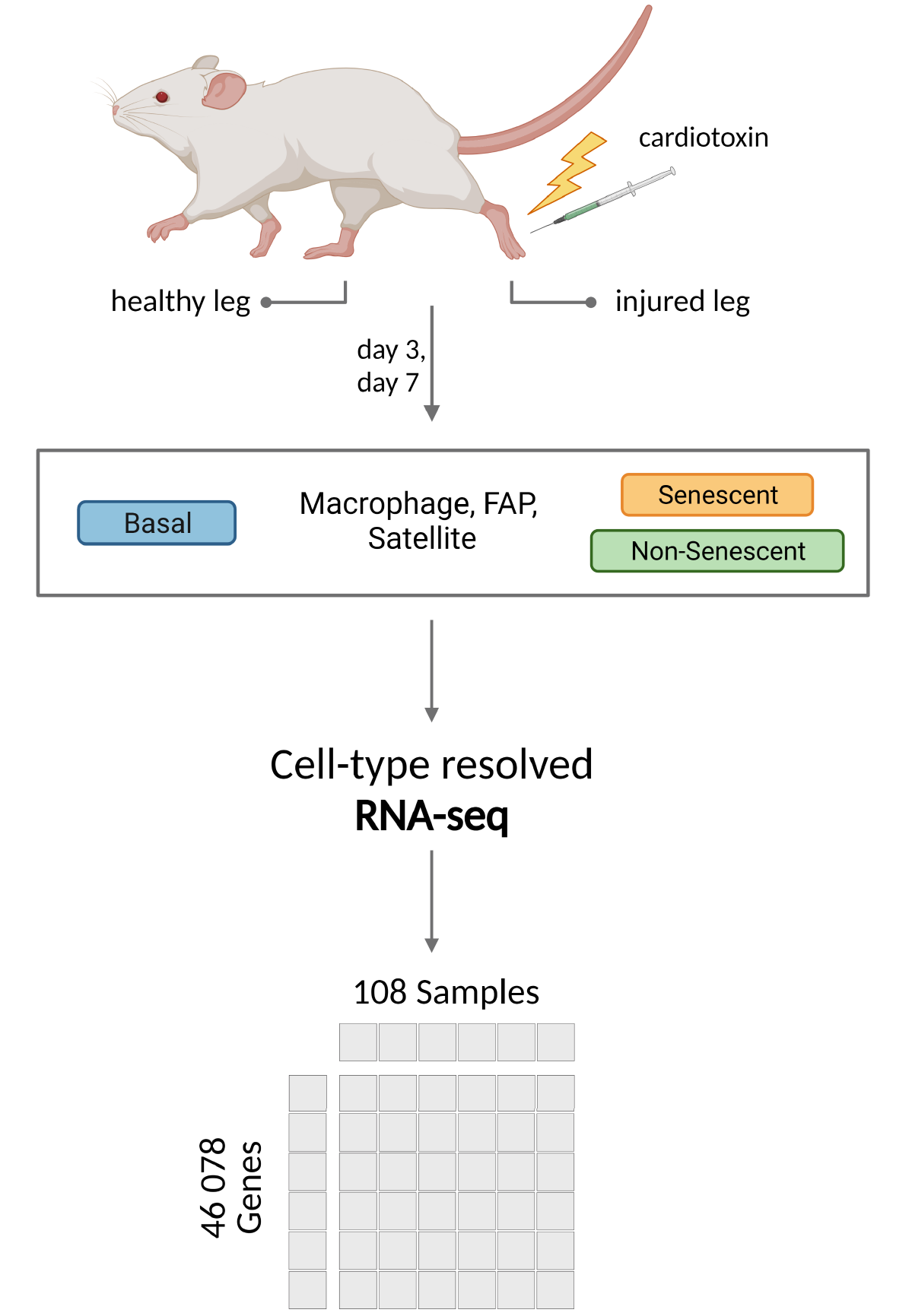}}
\caption{Workflow of the experimental design. Tissue is extracted from the injured and non-injured leg of young and geriatric mice at 3 and 7 days post-injury. 
Three cell types of interest are sorted out - macrophages, FAPs and satellite cells, each of them having three cell states: basal, senescent and non-senescent. The basal cells were extracted from the non-injured leg, while the senescent and non-senescent are taken from the injured one. 
A low input RNA sequencing protocol was performed to obtain the expression matrix, which consists of 108 samples each described by 46 078 genes.
}\label{Exp_design}
\end{figure}

Our study focuses on revealing the transcriptional regulation of cellular senescence during muscle repair and aging, by hypothesizing that cellular senescence can be discriminated from its gene expression profile. 
To that end, we analyzed the transcriptional profile obtained by bulk RNA sequencing of distinct cell types.
The cells were extracted from muscle tissue of healthy and injured mice, which were separated in three cellular states: senescent, non-senescent and basal (Fig.~\ref{Exp_design}). 

The data was obtained by performing experiments on mature adult mice belonging to two age groups -  young  (aged 3 months) and geriatric (aged 24 months). 
The experiments were conducted on a total of nine young and nine geriatric mice. 
The mice were subsequently divided into three pools after which the left leg was damaged with an injection of cardiotoxin, causing massive damage to the skeletal tissue \citep{garry2016}. 
Although the regenerative response happens almost immediately in both young and geriatric mice, senescent cells take a few days to emerge. 
The damaged muscle tissue was thus extracted 3 and 7 days after the injection.
When combined, these two measurements allow cellular differentiation within a heterogeneous population.
 The tissue was processed and the cell types were sorted as discussed in Sec.~\ref{sec:sorting} below.
In this study, we focus on three different cell types because of their crucial role in muscle tissue regeneration - satellite cells (muscle stem cells), fibro-adipogenic progenitor (FAP) cells, and macrophages \citep{joe2010,rigamonti2014}.
Each of the selected cell types was further categorized into three cell states: senescent and non-senescent, extracted from the injured (left) leg, and basal, extracted from the non-injured (right) leg. 
The RNA of the cells was collected and a low input RNA sequencing analysis was performed to determine the expression levels of a total of 46 078 genes, including protein-coding genes, non-coding genes, and pseudogenes. 

The study thus considered three cell types and three cell states in both young and geriatric mice, at two different time points (3 and 7 days after injury).
Since each condition was replicated three times, this resulted in a total of 108 possible combinations, which represented the 108 conditions considered in the analysis. 
In this context, each of the 108 conditions is characterized by a 46 078-dimensional vector, where each dimension corresponds to the expression level of a specific gene.
\section{Preliminary filtering}

\begin{figure}[htbp]
\centerline{\includegraphics[width=0.95\linewidth]{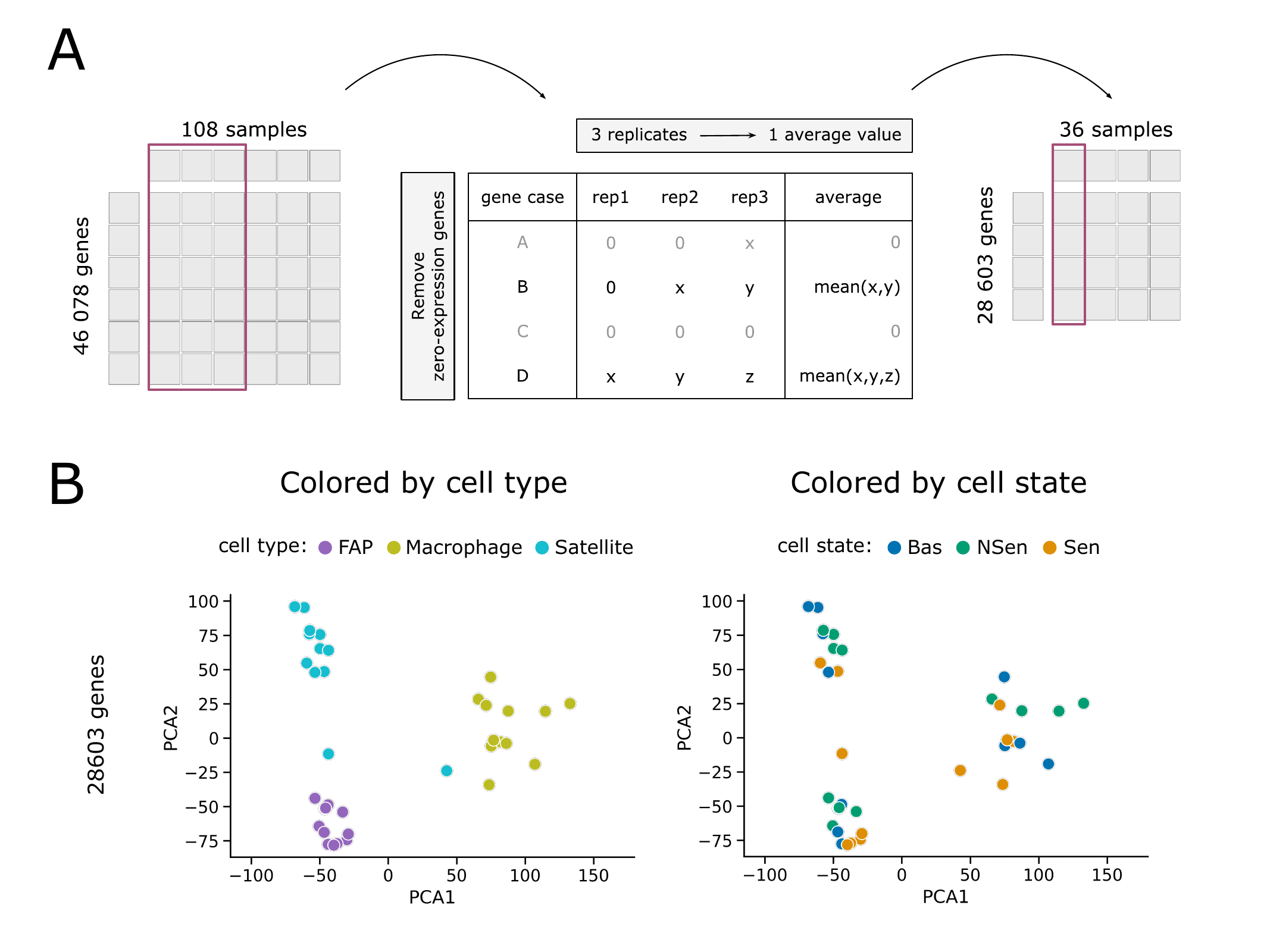}}
\caption{Dropout filtering. (A) The initial expression matrix consisted of 108 samples each defined by 46 078 genes. The 108 samples arise from 36 different cell conditions, each with 3 replicates. The first filtering consisted in averaging the expression of each gene among the 3 replicates of each condition and removing the ones whose expression was equal to 0 in at least 2 out of 3 replicates (gene cases A and C).
This combination of averaging and filtering led to an expression matrix consisting of 36 samples (representing the different cell conditions) and 28 603 genes.
(B) Principal component analysis of the remaining data, with cell conditions sorted either by cell type (left) or cell state (right).}
\label{fig:dropout_filtering}
\end{figure}

RNA sequencing is prone to exhibiting substantial technical noise, leading to dropout events in which a transcript is missing in a given replicate \cite{van2018recovering}.
To avoid this issue,  we averaged the expression of the three replicates for each condition and kept only the genes whose expression was non-zero in at least 2 out of 3 replicates. 
By doing so, we reduced the number of conditions to 36 and the number of genes to 28~603 (Fig.~\ref{fig:dropout_filtering}A).

We next asked to what extent the resulting filtering gene set was able to identify the senescent phenotype.
To that end, given the large size of the data set, we performed a dimensionality reduction via principal component analysis (PCA).
Figure~\ref{fig:dropout_filtering}B shows scatterplots of the different cell conditions for the first two principal components.
The conditions are color-coded depending on either the cell type (left panel) or cell state (right panel).
As can be seen in the plots, the 28~603 genes resulting from the filtering procedure described above are able to discriminate well between the three cell types analyzed (left panel), but not at all between cell states (right panel).
In other words, the selected genes are not sufficiently filtered to separate the senescent from the non-senescent phenotype.

\begin{figure}[htbp]
\centerline{\includegraphics[width=0.95\linewidth]{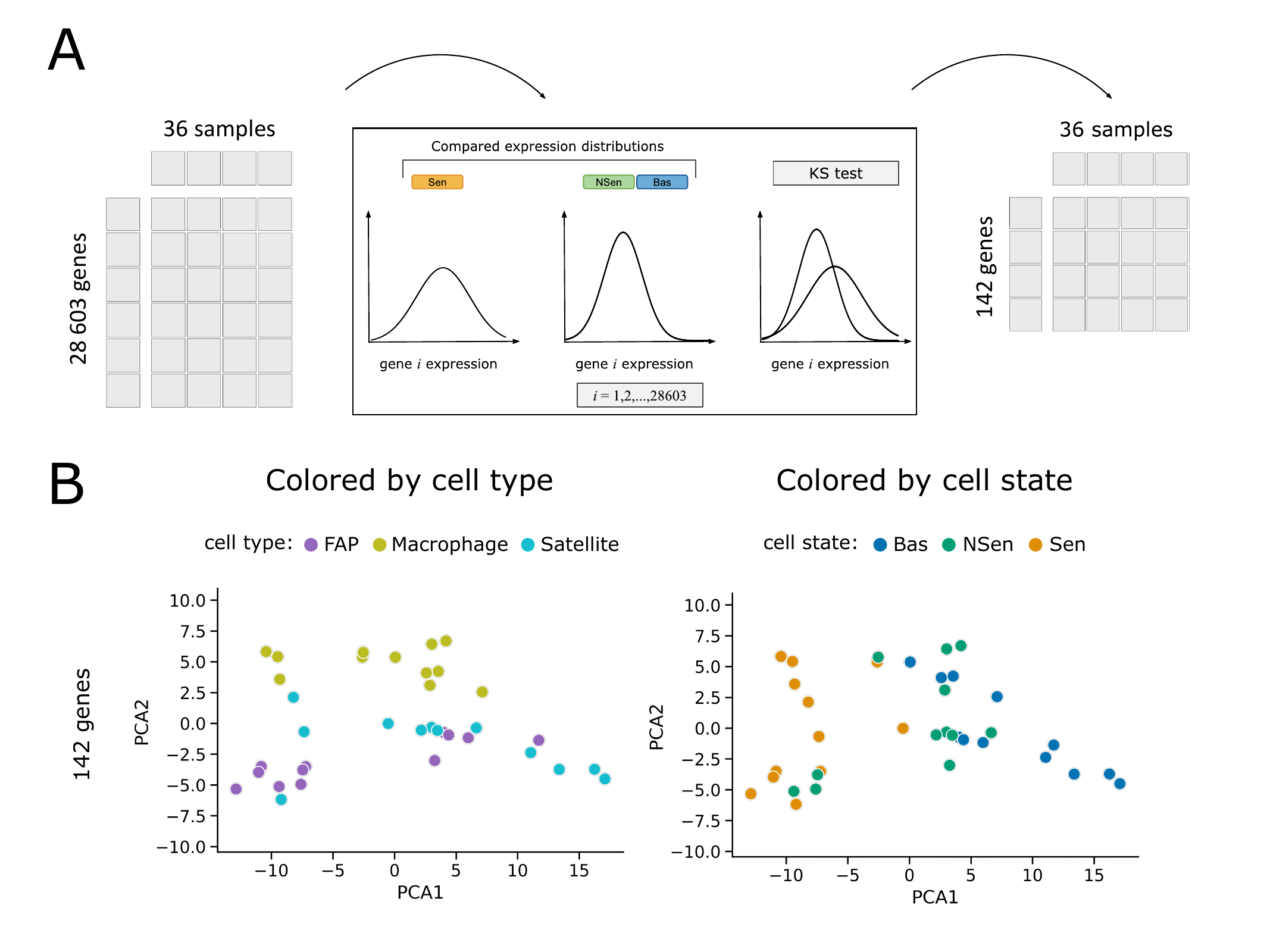}}
\caption{Significance filtering. (A) We compared the gene-expression distributions of the conditions between the senescent cell state and the non-senescent and basal states grouped together, for each one of the 28603 genes.
The distributions were compared by using the Kolmogorov-Smirnov test and applying a Bonferroni correction to the p-value of 0.05. A total of 142 genes passed the test.
(B) Principal component analysis of the remaining data, with cell conditions sorted either by cell type (left) or cell state (right).}
\label{fig:significance_filtering}
\end{figure}

Since our primary objective was to find the genes that can effectively discriminate between the three different cell states, rather than the cell types, further filtering was necessary. 
To that end, a Kolmogorov-Smirnov statistical test was conducted to compare the expression distributions of the senescent samples to the non-senescent and basal ones (Fig.~\ref{fig:significance_filtering}A).
After applying the Bonferroni statistical test adjustment to the p-value of 0.05 and eliminating the genes whose distributions were not significantly different among the cell states, we reduced the initial number of genes to 142.
The strong reduction in gene number eliminated the ability of the gene set to distinguish among cell types (Fig.~\ref{fig:significance_filtering}B, left panel), but
the data was now a bit more informative in distinguishing between cell states, as can be seen by comparing the right panels of Figs.~\ref{fig:dropout_filtering}B and \ref{fig:significance_filtering}B.

\section{Ranking and filtering genes via eigenvector centrality feature selection}

\begin{figure}[htbp]
\centerline{\includegraphics[width=0.95\linewidth]{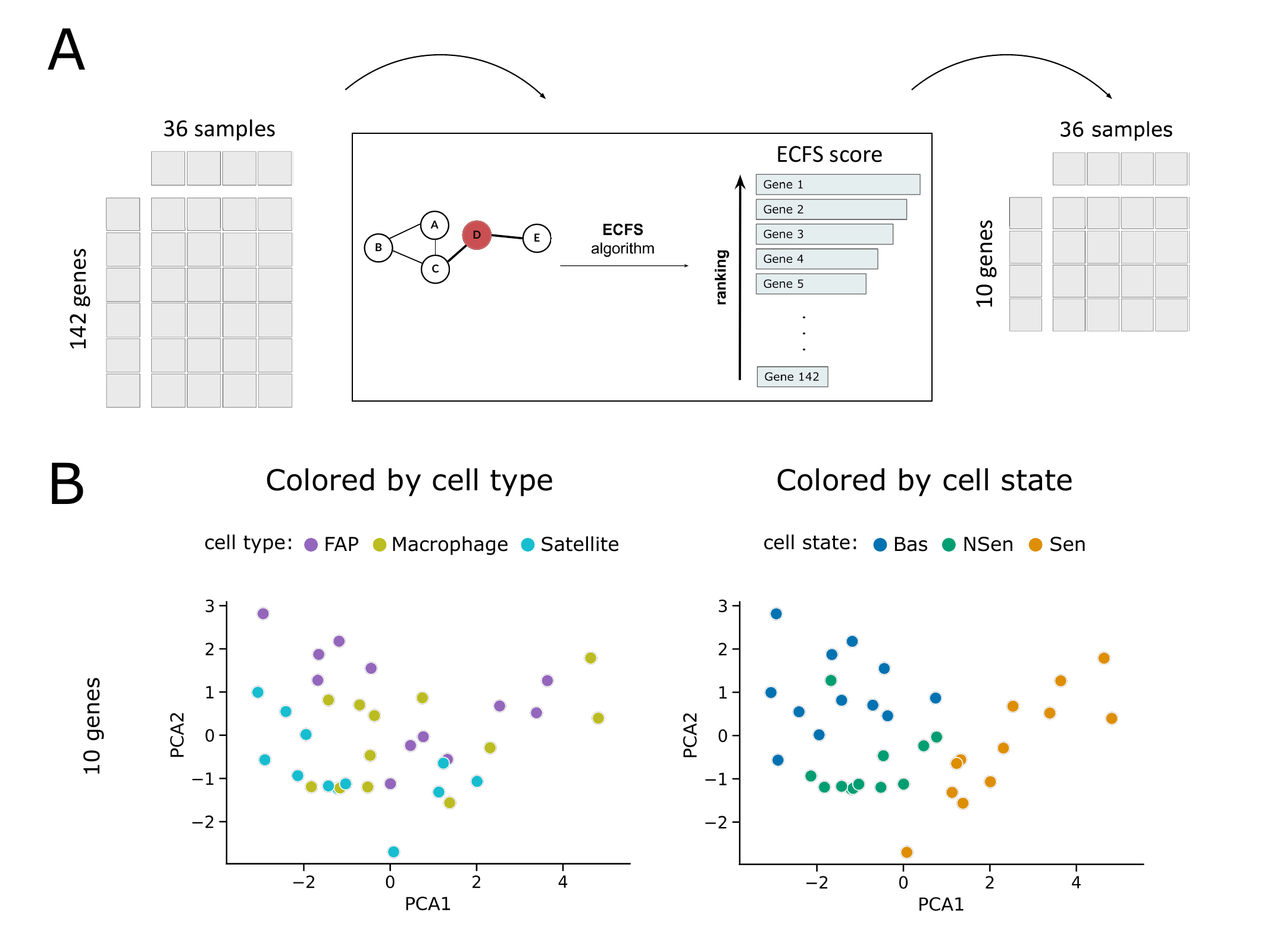}}
\caption{Ranking and filtering genes via eigenvector centrality feature selection. (A) The remaining 142 genes were ranked based on their importance in class separation by using the ECFS algorithm.
The top 10 ranked genes were retained, using the criteria discussed in Sec.~\ref{sec:netcomm}
below.
(B) Principal component analysis of the remaining data, with cell conditions sorted either by cell type (left) or cell state (right).}
\label{fig:ECFS_filtering}
\end{figure}

The preliminary filtering methods discussed above show that filtering the gene set enables a better separation between cell states (specifically the senescent versus non-senescent phenotypes) at the expense of the discrimination between cell types (which is not a priority in our case).
However the cell-state separation is still far from perfect.
We thus decided to focus further on the genes that are more strongly associated with the senescent phenotype.
To that end, we implemented the eigenvector centrality feature selection (ECFS) algorithm \citep{roffo2016} (Fig.~\ref{fig:ECFS_filtering}A).
The ECFS algorithm is a network-based method that aims to determine the most relevant features in a given classification task (in our case, we chose to distinguish between the non-senescent cells and the rest).
The algorithm uses a fully connected weighted network where the nodes are represented by the features.
Subsequently, it evaluates the discriminatory power of each feature by calculating the eigenvector centrality scores. 
In our case, the features are the genes, while the adjacency matrix is computed in terms of how discriminative each gene is for the stated separation. 
A detailed explanation of the use and implementation of the algorithm is given in Sec.~\ref{sec:ecfs} below. 

After applying the ECSF approach, we obtained a ranking of the 142 genes in order of importance for the chosen class separation. 
The higher the eigenvector centrality score, the more influential the gene is in differentiating between the various conditions.
This ranking allowed us to perform a final selection of the genes that we expect to have a key role in separating the senescent from the non-senescent phenotype, by enabling us to focus only on the highest ranked genes.
If we focus for instance on the 10 highest ranked genes, the PCA results show a clear improvement in the ability of those 10 genes (and those 10 genes only) in distinguishing the senescent phenotype from the rest.
The reason for choosing the top 10 genes is explained in the next Section.

\section{Network community analysis}
\label{sec:netcomm}

\begin{figure}[htbp]
\centerline{\includegraphics[width=0.75\linewidth]{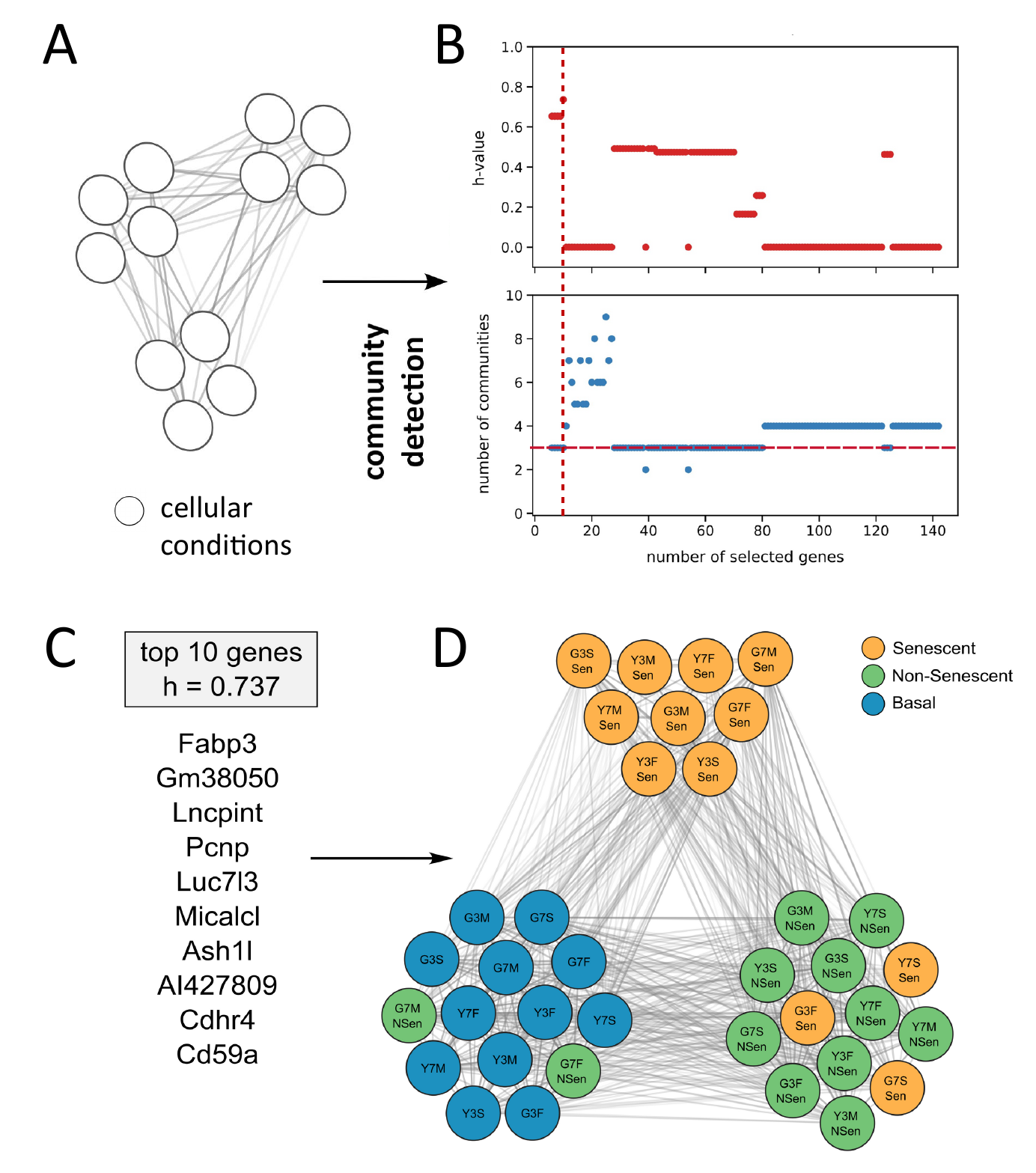}}
\caption{Network-community assessment. (A) Once the ECFS ranking was obtained, we proceeded to construct cell-state networks. The nodes of the networks are represented by the 36 cell conditions, while the edge weights correspond to the Pearson's correlation coefficients between the expression of the selected genes of the corresponding pair of conditions.
(B) ECFS ranking performance for increasing number of genes, when comparing the non-senescent phenotype with the rest.
The $h$ score and the number of communities are plotted in the top and bottom panels, respectively.
The best $h$-score of $0.737$ was obtained when considering the first 10 genes. The vertical and horizontal red dashed lines correspond to the optimal $h$ score and the three communities, respectively.
(C,D) For that case we obtain the best community separation, represented here.
The genes listed on panel C give rise to 3 communities, each one predominantly consisting of a single cell state --basal, senescent or non-senescent--, each represented in panel D by a different color.
The node labels have the format AnB, where A is either Y or G, corresponding to the sample coming from a young or geriatric mouse, respectivel, n is either 3 or 7, corresponding to a sample obtained 3 or 7 days after the injury, and B is either M, F or S, corresponding to the cell type being a macrophage, FAP, or satellite cell, respectively.}
\label{fig:communities}
\end{figure}

After ranking the genes with the ECFS algorithm, we proceeded by building network considering an increasing number $N$ of ranked genes, adding one gene at a time in the order of the ranking in each iteration of the method. 
The nodes of this network correspond to the 36 different cell conditions, while the edges are established by an adjacency matrix based on the Pearson's correlation coefficients between the $N$ selected genes for each pair of conditions (Fig.~\ref{fig:communities}A).
Once the network was constructed, we proceeded to find the network communities, using the Louvain algorithm for community detection \citep{blondel2008}. 
The process was repeated until all 142 genes were included.
The number of communities obtained for increasing number of ranked genes included in the network is shown in the bottom panel of Fig.~\ref{fig:communities}B.

Ideally, the community detection algorithm should identify exactly three communities, with each cell state (senescent, non-senescent, and basal) belonging to its own individual community. 
Additionally, it was important to quantify the separation quality of the communities, i.e. how well the communities obtained with a given selection of genes discriminate between the cell states. 
To do so, we introduced a goodness of separation measure based on Shannon's entropy $H = -\sum_{i} p_{i}\ln(p_{i})$, where $p_{i}$ is the probability of occurrence of a specific event, in this case the fraction of cell conditions belonging to a specific community.
The entropy was computed with respect to both communities and cell states, and the total entropy $H_{\rm tot}$ of a given community separation was obtained by summing the individual entropies.
For a thorough description of this process along with an illustrative example, see Sec.~\ref{sec:sep} below. 

Once we have computed the total entropy of the community separation, we now define the goodness of separation by:
\begin{align}
h = \begin{cases}
    \dfrac{H_{\rm max}- H_{\rm tot}}{H_{\rm max}}, & \text{if } n = 3 \\
    0, & \text{if } n \neq 3,
\end{cases}
\end{align}
where $n$ is the number of communities found by the community detection algorithm, and $H_{\rm max}$ is the maximum possible value of the entropy, which corresponds to the case when there is no clustering and each cell condition is in its own community (36 communities in total).
Furthermore, the total entropy $H_{\rm tot}$ is normalized by $H_{\rm max}$, and its sign is inverted so that an $h$-value of 0 indicates maximum disorder, while a value of 1 indicates perfect separability of the three cell states, each belonging to its own community (see Sec.~\ref{sec:sep}).
Finally, we consider for assessment only the cases where exactly three communities were detected.
The values of the $h$ score for increasing number of genes is shown in the top panel of Fig.~\ref{fig:communities}B. 

As shown in Fig.~\ref{fig:communities}B, the best cell state separation was obtained when the top 10 ranked genes from the ECFS algorithm output were considered.
The corresponding genes are listed in Fig.~\ref{fig:communities}C, and the corresponding network is shown in Fig.~\ref{fig:communities}D, with the different cell states shown by different colors.
As it can be observed, one community contains 9 out of 12 senescent cells without containing any other cell state.
Additionally, the community primarily composed of basal cells includes two conditions that belong to the non-senescent state.
This may be explained by the fact that non-senescent cells resemble basal cells, as they have not yet developed any traits specific to senescence by the time the tissue was collected.
The predominantly non-senescent community contains three samples belonging to the senescent state. 
Similarly to the mixing of basal and non-senescent cells, this might be because both of these cell states are extracted from the injured leg.

To test the robustness of the obtained separation, we performed a permutation test that consisted of randomly changing the labels of the 36 cell conditions 1000 times.
Those artificial datasets then underwent the ECFS algorithm, and the community detection was applied on a network constructed with the 10 top-ranking genes. 
We found that it is significantly unlikely that such separation could be obtained by chance (p-value = $0.0019$).
This analysis is discussed in Sec.~\ref{sec:stat}, and the corresponding distribution of randomly computed normalized entropy scores $h$ is given in Figure \ref{Histogram_h_scores}.

\section{Discussion}

A final filtering of the gene list was performed by removing each one of the 10 genes obtained above, and determining which ones of these removals resulted in a noticeable disruption of the community structure, resulting in a modification of the separation of cell states.
The results of this analysis are shown in Table~\ref{tab:removal}.
Furthermore, if only the 6 highlighted genes are considered for network construction and community detection, they produce the same separation among cell states as the 10 best-ranked genes.

\begin{table}[htbp]
   \centering
   \begin{tabular}{|c|c|} 
\hline
\textbf{Eliminated gene} & \textbf{$h$ score} \\
\hline
\hline
\underline{Fabp3} & {0.605} \\
Gm38050 & 0.737 \\
\underline{Lncpint} & {0} \\
\underline{Pcnp} & {0} \\
\underline{Luc7l3} & {0} \\
Micalcl & 0.737 \\
Ash1l &0.737 \\
\underline{Al427809} & {0} \\
Cdhr4 & 0.737 \\
\underline{Cd59a} & {0.654} \\
\hline
   \end{tabular}
   \caption{Effect of eliminating one by one each gene on the goodness of separation.
   Those genes whose absence noticeably disrupts the cell-state separation are underlined.}
   \label{tab:removal}
\end{table}

Some of the 6 genes highlighted in Table~\ref{tab:removal} were found to be directly associated with senescence.
For instance, one of the genes most relevant for the separation was Lncpint, a p53-induced long intergenic non-coding transcript.
Removing it from the analysis resulted in a complete disruption of the otherwise well-defined clusters. 
In that case, there was no separation of senescent cells whatsoever - indicating the crucial role of this transcript in the differentiation of this cell state. 
Recently, there have been numerous reports about the role of Lncpint in various biological processes ranging from DNA damage responses \citep{wang2021}, cell cycle and growth arrest \citep{bukhari2022}, cellular senescence \citep{xiang2021}, cell migration \citep{he2021} and apoptosis \citep{bukhari2022}. 
These findings are consistent with our results.
A second of the highlighted genes, Fabp3 (Fatty acid-binding protein 3), is involved in the intracellular transport of long-chain fatty acids.
Fabp3 has been reported to be upregulated in senescent cells, together with numerous other lipid-transport and lipid metabolism genes \citep{moiseeva2023}.
In general, lipid uptake plays a role among the generally accepted hallmarks of senescence, which
include cell cycle arrest, resistance of apoptosis, metabolic and morphological changes, and highly secretory phenotype (SASP) \citep{gonzalez-gualda2021}.
Lipids are found to be essential for each of these features \citep{hamsanathan2022}.
Furthermore, an accumulation of lipid droplets in senescent cells has been reported in various studies \citep{lizardo2017, flor2017, ogrodnik2019}.
%

\begin{figure}[htbp] 
\centerline{\includegraphics[width=0.8\linewidth]{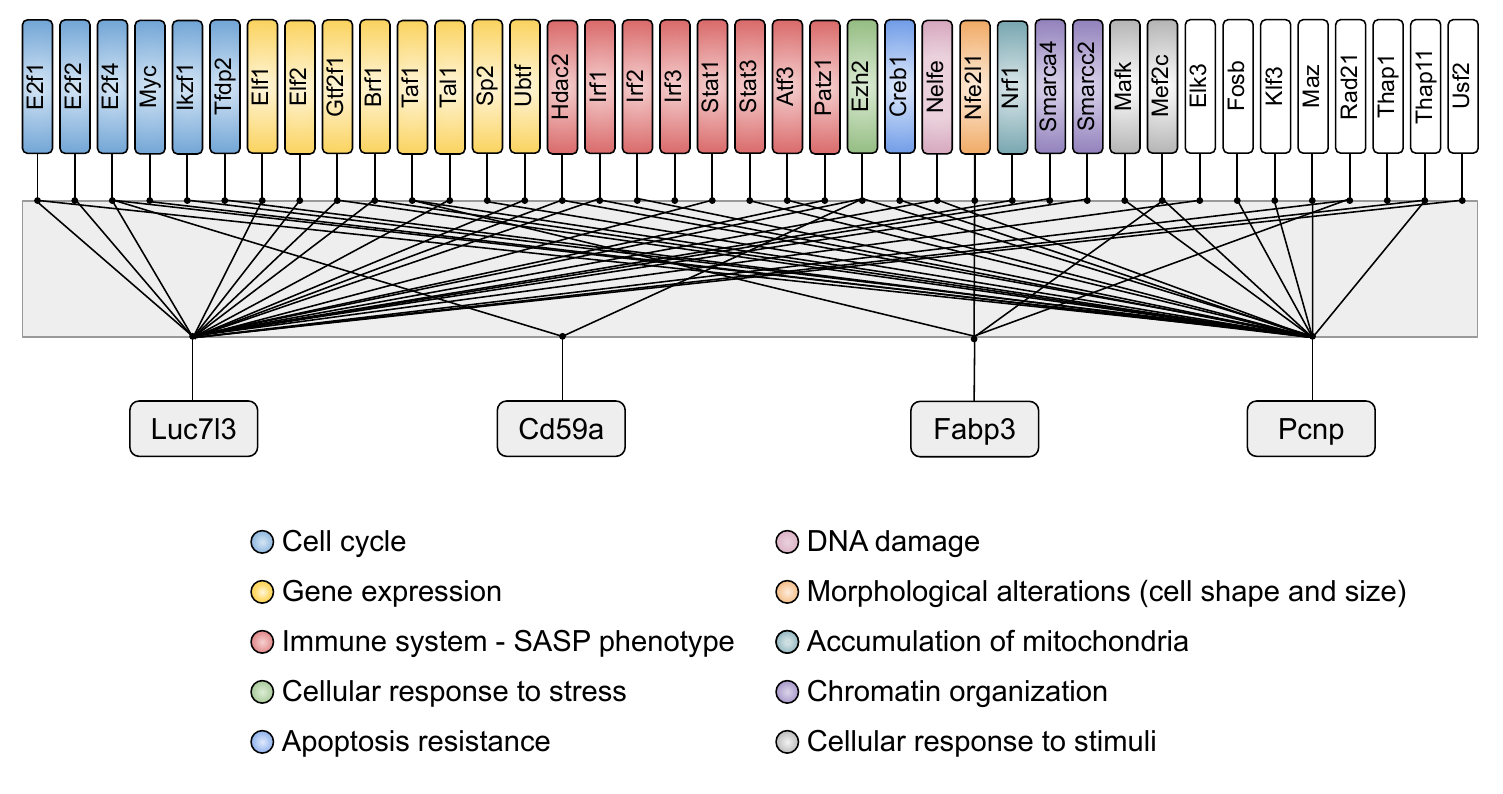}}
\caption{Transcription factors of the found genes and their role in senescence. The transcription factors from the most relevant protein-coding genes are shown grouped by color, representing the pathways they are involved in. The shown pathways are related to the established hallmarks of senescence.}
\label{Molecular_hallmarks_of_senescence}
\end{figure}

To provide an additional assessment of the other genes involved in the cell state separation, we conducted an upstream analysis of their transcription factors (TFs).
Specifically, out of the 6 essential genes found to separate the clusters, 4 were identified as protein-coding.
The transcription factors regulating those genes were extracted from the Gene Regulatory Network database (GRNdb) \citep{fang2020}. 
Given that the analyzed samples were extracted from muscle tissue, we used the TF-target regulations for mouse muscle available in the database.
Subsequently, the implication of a particular transcription factor in a pathway was assessed using the KEGG Pathways database \citep{kanehisa2000}.
Finally, the transcription factors were then grouped according to their involvement in the pathways associated with the established hallmarks of senescence.    
As shown in Figure \ref{Molecular_hallmarks_of_senescence}, many of the identified transcription factors were involved in pathways linked to several hallmarks of senescence such as cell cycle, secretory phenotype, cellular response to stress, apoptosis resistance, DNA damage, morphological alterations, accumulation of mitochondria, chromatin organization and cellular response to stimuli.

\begin{figure}[!ht]
\centerline{\includegraphics[width=0.6\linewidth]{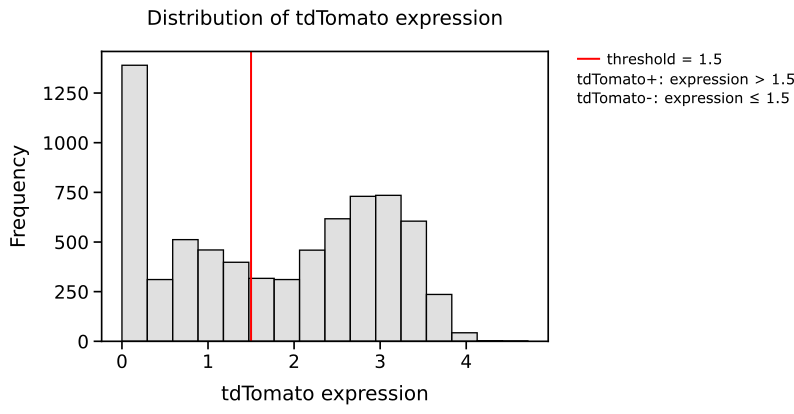}}
\caption{Distribution of p16 expression, as measure by the fluorescent marker tdTomato.
The expression among all the considered cell types presents a bimodal distribution.
The threshold for establishing p16+ and p16- cells is chosen to be 1.5.}
\label{tdTomato_distribution}
\end{figure}

Finally, to validate our findings further, we analyzed a previously published single-cell RNA-seq dataset in normal and fibrotic liver and kidney tissue \cite{omori2020}.
This study used the levels of p16 as a marker of senescence, since senescent cells are known to have an increased expression of that protein \citep{rayess2012},
First, we split the population into two subgroups based on the expression on the fluorescent marker tdTomato, which represents the expression of p16 in this dataset.
The expression distribution of tdTomato, together with the chosen expression threshold, is shown in Figure \ref{tdTomato_distribution}.

\begin{figure}[htbp] 
\centerline{\includegraphics[width=1\linewidth]{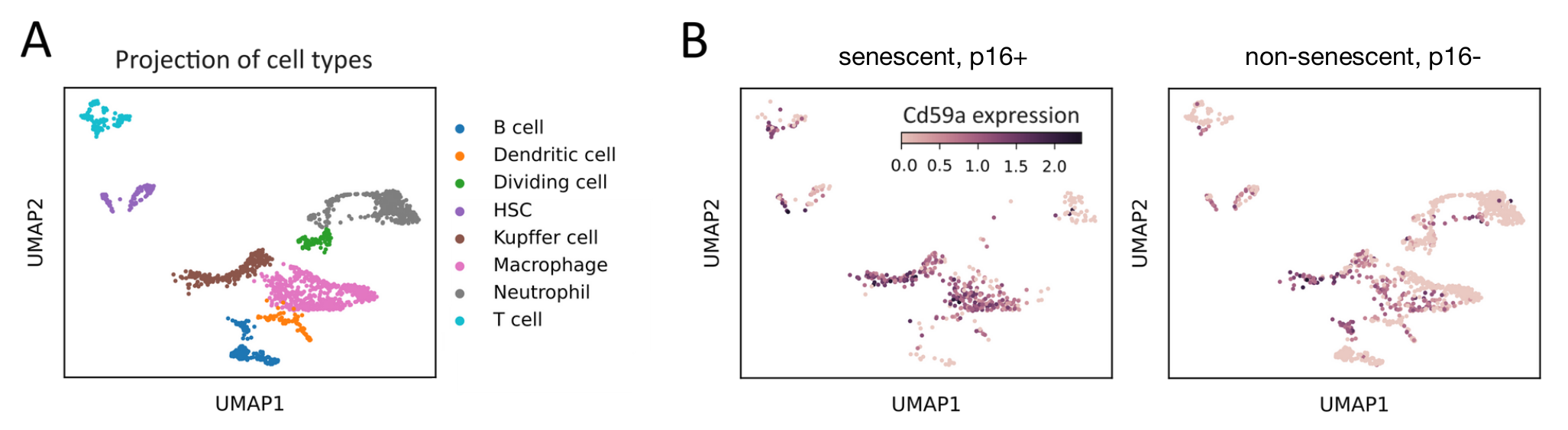}}
\caption{Expression of Cd59a in p16+ and p16- cells. (a) The UMAP projection of the cell types. (b) The cells are classified into two groups based on their tdTomato expression. Cells with tdTomato expression larger than 1.5 are considered tdTomato+, whereas cells that have the expression of tdTomato $\leq 1.5$ are labeled as tdTomato-. A predominant presence of increased Cd59a expression is observed in the tdTomato+ cells, indicating that Cd59a could be a possible marker of senescence for the shown cell types.}
\label{tdTomato_Cd59a_expression}
\end{figure}

\begin{figure}[htbp]
\centerline{\includegraphics[width=1\linewidth]{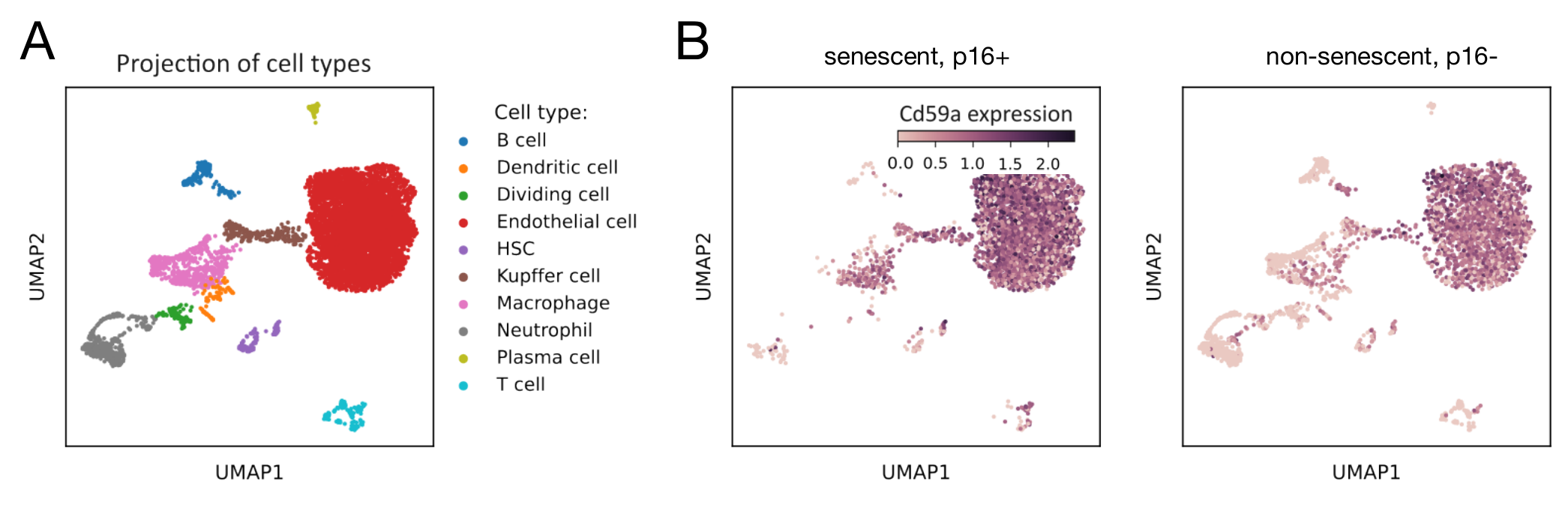}}
\caption{Expression of Cd59a in tdTomato+ and tdTomato- cells. (a) The UMAP projection of all the considered cell types. (b) The cells are classified into two groups based on their tdTomato expression. Cells with tdTomato expression larger than 1.5 are considered tdTomato+, whereas cells that have the expression of tdTomato $\leq 1.5$ are labeled as tdTomato-. The cells in the tdTomato+ panel show a higher expression of C59a, except for the endothelial and plasma cells.}
\label{tdTomato_Cd59a_expression_all_cells}
\end{figure}

The analysis above allowed us to establish a ground-truth separation between senescent (p16+, with tdTomato expression above the threshold) and non-senescent cells (p16-, with tdTomato expression below the threshold).
The cells were project in a low dimensional UMAP representation of their transcriptome for the two cell states, as shown in Fig.~\ref{tdTomato_Cd59a_expression}A.
We then superimposed the expression of one of the 6 genes identified with our analysis, Cd59a (which had not been associated so far with senescence) on this UMAP representation.
As can be seen when comparing the two panels of Fig.~\ref{tdTomato_Cd59a_expression}B, Cd59a exhibited a clear enrichment in the senescent cell populations, larger than other traditional markers.
The best categorization can be observed among Kupffer cells and macrophages, where the alignment between a cell being p16 positive and expressing a high level of Cd59a is the highest.
The cells that showed the least matching between tdTomato+ and Cd59a were the endothelial and plasma cells, shown together with the other cell types in Fig.~\ref{tdTomato_Cd59a_expression_all_cells}. 
It is worth noting that the cell types used for validation originate from an entirely distinct tissue from the ones employed in our analysis, which highlights the potential generalizability of our conclusions.

\section{Conclusions}
In order to better understand the transcriptional regulation landscape of cellular senescence in the tissue regeneration process, we conducted an analysis that leans on network-theory concepts such as community detection and eigenvector centrality. We found that, from a transcriptional point of view, senescence is a heterogeneous process characterized by a variety of molecular hallmarks. In search of these hallmarks, we identified a gene set that underlies the senescent phenotype. The relevance of the identified genes was assessed in terms of their transcriptional roles, which were found to correspond to several established phenotypical hallmarks of senescence.

Our findings aim to shed light on the transcriptional regulation landscape of cellular senescence in tissue regeneration and potentially contribute to the understanding of this heterogeneous process.
Specifically, our analysis resulted in a set of 6 genes that are able to separate almost completely the senescent phenotype from the non-senescent and basal ones. 
Despite observing some mixing within the found clusters, we assume it could be attributed to certain shared characteristics.
For instance, some non-senescent cells clustered together with the basal cells - probably because both cell types exhibit the same non-senescent features, even though they were extracted from different legs of the mice.
Furthermore, some senescent cells were clustering with the non-senescent community. 
Since both of those cells were extracted from the same damaged leg of the mice, it is expected that they might share some transcriptional similarity.
However, it is worth noting that there are no senescent cells present in the community consisting of predominantly basal cells and vice versa.
This result supports the hypothesis that basal and senescent cells have a sufficiently different transcriptional signature.
The genes found by our analysis might be associated with the already established characteristics of senescent cells and could, therefore, facilitate the identification of the molecular hallmarks of senescence.

\section{Materials and methods}

\subsection{Cell sorting}
\label{sec:sorting}

After the filtering, digestion and purification of the extracted tissue, cell sorting by flow cytometry was performed to isolate the cell types of interest. 
Precisely, cells were selected by using forward- and side-scatter detectors.
The forward-scatter measurement allows discriminating cells by size while measuring the side scatter provides information about the granularity of a cell.
When combined, these two measurements allow cellular differentiation within a heterogeneous population.
The entire experimental protocol, together with the RNA extraction procedure, was done as described by Moiseeva et al.\cite{moiseeva2023}.

\subsection{Eigenvector centrality feature selection}
\label{sec:ecfs}
The eigenvector centrality feature selection (ECFS) algorithm \citep{roffo2016} is based on mapping the given feature selection (FS) problem to a graph in which the nodes represent the features of interest. 
In our analysis, the features are the genes that separate the predetermined classes of cell states. 
Specifically, given a set of features $X = \{x^{(1)},...,x^{(n)}\}$, a fully-connected undirected graph $G = (V, E)$ is built. 
$V$ represents a set of vertices corresponding to each feature $x$, whereas $E$ represents the weighted edges among those features. 
The weighted edges are defined by the adjacency matrix $A$ that consists of two measures, one considering the relevance of a given feature in a supervised way and the other including redundancy in an unsupervised way.
The first measure considered to compute the matrix is the Fisher criterion, defined as:
\begin{align}
f_{i}= \frac{\vert \mu_{i,1} -  \mu_{i,2}\vert ^{2}}{\sigma_{i,1}^{2}+\sigma_{i,2}^{2}},
\end{align}
where $\mu_{i,C}$ and $\mu_{i,C}$ represent the mean and the standard deviation of the $i$-th feature when considering the samples of the $C$-th group. 
A feature is considered to be more discriminative the higher the value of $f_{i}$. 
Next, mutual information is used to assign a high score to the features related to the given classes:
\begin{align}
m_{i}= \sum_{y \in Y} \sum_{z \in x^{(i)}}  p(z,y) \log \left( \frac{p(z,y)}{p(z)p(y)}\right),
\end{align}
where $Y$ represents the set of class labels and $p(\cdot, \cdot)$ the joint probability distribution.
The kernel $k$ for the adjacency matrix $A$ is then obtained by the following matrix product:
\begin{align}
k = \left( f \cdot m^{\top}\right)
\end{align}
The kernel $k$ accounts for the supervised part of the measure, given that it is computed taking into account the class label.

The second metric, defined in an unsupervised way, is based on standard deviation. Precisely, it captures the amount of dispersion of the features from the average:
\begin{align}
\Sigma(i,j)= \left( \sigma^{(i)},\sigma^{(j)} \right),
\end{align}
Finally, the adjacency matrix of the graph $G$ taking into account both metrics is given by:
\begin{align}
A= \alpha k + (1-\alpha)\Sigma,
\end{align}
where $\alpha$ is the loading coefficient, which should be chosen to fit the purposes of the feature selection. 
When computing the feature selection ranking for our set of genes, we consider only the supervised part of the matrix, given that we know the labels of our classes (senescent, non-senescent and basal).
Therefore, by putting $\alpha = 1$, the proposed adjacency matrix is simplified to $A = k$.
Once the adjacency matrix $A$ is constructed, the network is fully defined. 

The feature selection process relies on finding the eigenvector centrality for the defined network.
The eigenvector centrality measures the influence of a node in a network.
Therefore, a node would have a high centrality score if it is connected to many nodes or in the case it is connected to highly influential nodes \citep{bonacich1972}. 
Let $G = (V, E)$ be a graph and $A$ its adjacency matrix with a generic entry $a_{ij}$.
The relative centrality score of a node $x_{v}$ is proportional to the sum of the centralities of the nodes to which it is connected.
Hence,
\begin{align}
x_{v} = \frac{1}{\lambda} \sum_{j=1}^{n}a_{ij}x_{j}, \quad i = 1,...,n
\end{align}
where $n$ is the number of nodes in the network and $\lambda$ is a constant.
By reorganizing the previous equation, we obtain the eigenvector equation:
\begin{align}
A \textbf{x} = \lambda \textbf{x}.
\end{align}
In general, there are many different eigenvalues $\lambda$ that satisfy the previous equation.
However, in the case of non-negative square matrices, the Perron-Frobenius theorem guarantees the existence of a positive real eigenvalue that is larger than all other eigenvalues, and its corresponding eigenvector is strictly positive (all components are positive).
Therefore, the $\lambda$ value of interest is the largest eigenvalue of matrix $A$.
The $v^{th}$ component of the related eigenvector results in the desired centrality score, which corresponds to the ranking of the $v^{th}$ feature in the ECFS algorithm.
%
%
%
\subsection{Measuring the goodness of separation}
\label{sec:sep}

\begin{figure}[htbp]
\centerline{\includegraphics[width=0.8\linewidth]{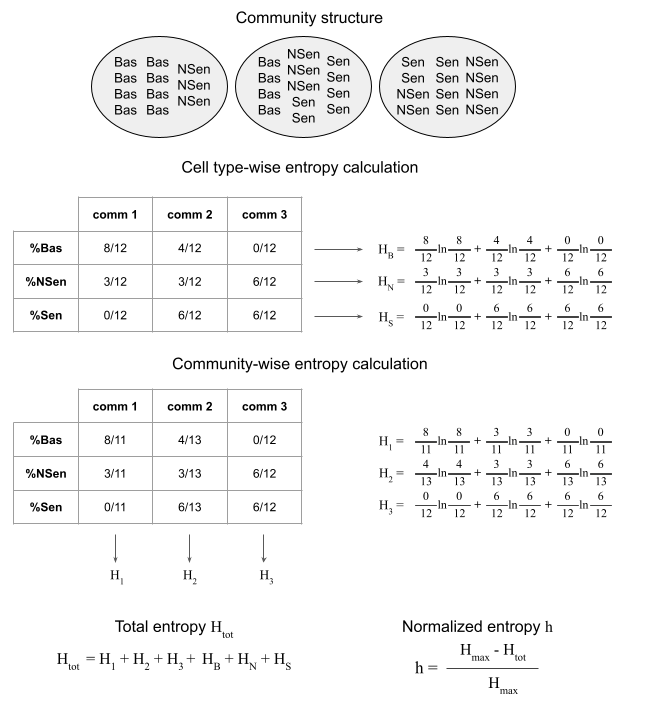}}
\caption{Entropy calculation example. 
To obtain the total entropy $H_{\rm tot}$, first, we have to calculate the individual entropies, both cell type-wise and community-wise. Since a total of 12 senescent, 12 non-senescent and 12 basal cells were present in our analysis, in the cell type-wise entropy computation, the probability $p_{i}$ is defined as the probability of a cell found in a community to belong to a given cell state. 
In the community-wise entropy calculation, the probability $p_{i}$ is defined by the ratio of cells belonging to a specific cell state to the overall number of cells within that community.
The total entropy $H_{\rm tot}$ is calculated by summing all the individual cell type-wise and community-wise entropies. 
The normalized entropy value $h$ is obtained by normalizing the total entropy $H_{\rm tot}$ with the maximum possible entropy $H_{\rm max}$ that corresponds to the case when each cell belongs to its own separate community.}
\label{Entropy_calculation_example}
\end{figure}

Since the data of our analysis consists of three different cell states (basal, senescent and non-senescent), the goodness of separation score is calculated only in the cases when the number of communities $n$ detected by the community detection algorithm is exactly 3. 
Figure~\ref{Entropy_calculation_example} provides an illustrative example of how the goodness of separation is determined.
First, the probability of occurrence of a given cell state $p_{i}$ was defined for every cell state. 
Taking into account that the 36 analyzed conditions include 12 senescent, 12 non-senescent, and 12 basal cells, in the cell type-wise entropy calculation, each entry $p_{i}$ corresponds to the probability of a cell present in a community to belong to a given cell state.
Therefore, when summing the values of $p_{i}$ for each row, a total probability of 1 is obtained (first table in Fig.~\ref{Entropy_calculation_example}).

In the community-wise calculation, each entry $p_{i}$ represents the ratio of cells belonging to a cell state and the total number of cells in that community. 
In this case, the total probability of 1 is achieved by summing up the entries $p_{i}$ for each column, as shown in the lower table in Figure \ref{Entropy_calculation_example}.
Subsequently, Shannon's entropy defined as $H = -\sum_{i} p_{i}\ln(p_{i})$ is calculated for each of the cell states and cell types.
The total entropy $H_{\rm tot}$ is defined as the sum of the individual entropies.
To keep the entropy values in the interval between 0 and 1, a normalized entropy value $h$ is computed. 
The normalization was done by taking into account the maximum possible entropy $H_{\rm max}$ that corresponds to the case in which each cell belongs to its own community. 
The normalized entropy value is given as:
\begin{align}
h = \begin{cases}
    \dfrac{H_{\rm max}- H_{\rm tot}}{H_{\rm max}}, & \text{if } n = 3 \\
    0, & \text{if } n \neq 3,
\end{cases}
\end{align}
where $n$ is the number of communities found by the community detection algorithm and $H_{\rm max} = -3\ln\frac{1}{12}$.
%

%
\subsection{Statistical validation}
\label{sec:stat}

\begin{figure}[!h]
\centerline{\includegraphics[width=0.6\linewidth]{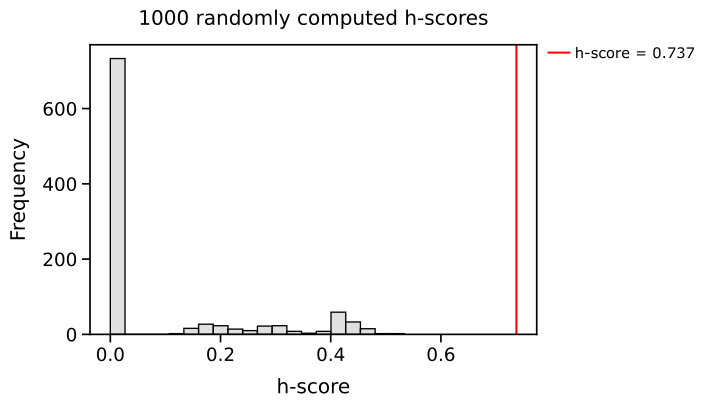}}
\caption{Histogram of randomly computed $h$-scores. The $h$-scores represented in the histogram were computed by permuting the cell conditions labels.  The artificially created labels were sorted into 2 classes, one consisting of non-senescent cells and the other of senescent and basal cells. The ECFS algorithm was performed for the classes, and the corresponding networks with the communities were computed. The $h$-score obtained for the 10 best-performing genes was extracted. The described procedure was repeated 1000 times and the corresponding $h$-scores are shown in the histogram. The red line represents the original maximum $h$-score obtained by our separation ($h = 0.737$). The p-value for our obtained $h$-score to be part of this distribution was found to be 0.00019.}
\label{Histogram_h_scores}
\end{figure}

To assess the robustness of the separation obtained by the 10 selected genes, a randomized permutation test was performed. 
The test consisted in generating 1000 datasets by permuting the labels of the 36 cell conditions.
Each dataset was first put through the ECFS algorithm to obtain the ranking of the genes, after which we proceeded with the network formation and community detection for the top 10 ranked genes.
By doing so, we obtained a distribution of the calculated normalized entropy scores $h$. 
The normalized entropy score from the original dataset $h$ was then compared to the distribution by a Kolmogorov-Smirnov test.
The test results showed that it is significantly unlikely to obtain such separation by chance, giving a p-value of $0.0019$.
The distribution of $h$-scores together with the $h$-score from the original dataset are shown in Figure \ref{Histogram_h_scores}.

\section*{Acknowledgments}

This work was supported by project PID2021-127311NB-I00 financed by the Spanish Ministry of Science and Innovation, the Spanish State Research Agency and FEDER (reference MICIN/AEI/10.13039/501100011033/FEDER), by the Maria de Maeztu Programme for Units of Excellence in R\&D (project CEX2018-000792-M), and by the Generalitat de Catalunya (ICREA Academia programme).

\end{document}